\documentclass[pre,onecolumn,superscriptaddress,groupedaddress]{revtex4}
\newcommand {\fptd} {\rho_B(t\,|\,x_0)} 
\newcommand {\fptdn} {\rho_B(n\,|\,x_0)}

\usepackage{float}
\usepackage{graphicx}
\usepackage{color}
\usepackage{amssymb}
\usepackage{amsmath}

\begin{document}

%%%%%%%%%%%%%%%%%%%%%%%%%%%%%%%%%%%%%%
%
%	TITLE, AUTHORS, ETC...
%
%%%%%%%%%%%%%%%%%%%%%%%%%%%%%%%%%%%%%%
%\title{First--Passage  of a Brownian Particle -- Escape from a Bounded Potential }
\title{A simple method to calculate first--passage time  densities of non--smooth processes}
%\author{Markus Nyberg, Tobias Ambj\"{o}rnsson and Ludvig Lizana}
\date{\today}

\author{Markus Nyberg} \email{markus.nyberg@umu.se} \affiliation{Integrated Science Lab, Department of Physics, Ume\r{a} University, SE-901 87 Ume\r{a}, Sweden}
\author{Tobias Ambj\"{o}rnsson} \affiliation{Department of Astronomy and Theoretical Physics, Lund University, S\"{o}lvegatan 14A, SE-223 62 Lund, Sweden}
\author{Ludvig Lizana} \affiliation{Integrated Science Lab, Department of Physics, Ume\r{a} University, SE-901 87 Ume\r{a}, Sweden }

%\author{Markus Nyberg}
%\email{markus.nyberg@umu.se}
%\affiliation{Integrated Science Lab, Department of Physics, Ume\r{a} University, SE-901 87 Ume\r{a}, Sweden}
%\author{Tobias Ambj\"{o}rnsson}
%\affiliation{Department of Astronomy and Theoretical Physics, Lund University, S\"{o}lvegatan 14A, SE-223 62 Lund, Sweden}
%\author{Ludvig Lizana}
%\affiliation{Integrated Science Lab, Department of Physics, Ume\r{a} University, SE-901 87 Ume\r{a}, Sweden }

%%%%%%%%%%%%%%%%%%%%%%%%%%%%%%%%%%%%%%
%
%	ABSTRACT
%
%%%%%%%%%%%%%%%%%%%%%%%%%%%%%%%%%%%%%%

\begin{abstract}
%\begin{center}
Numerous applications all the way from biology and physics to economics depend on the density of first crossings over a boundary. Motivated by the lack of analytical tools for computing first--passage time densities (FPTDs) for complex problems, we propose a new simple method based on the Independent Interval Approximation (IIA). We generalise previous formulations of the IIA to handle non--smooth processes,  and derive a closed form expression for the FPTD in Laplace and $z$--transform space for arbitrary boundary and starting points in one dimension. We focus on Markov processes for which the IIA is exact. To apply our equations, we calculate the FPTD in two cases: the Ornstein--Uhlenbeck process and the discrete time Brownian walk.  Our results are in good agreement with Langevin dynamics simulations. 
%We anticipate that our result will have a wide applicability in a number of escape problems.
%\end{center}
\end{abstract}

\maketitle

%%%%%%%%%%%%%%%%%%%%%%%%%%%%%%%%%%%%%%
%
%	I N T R O D U C T I O N
%
%%%%%%%%%%%%%%%%%%%%%%%%%%%%%%%%%%%%%%

\section{Introduction} \label{sec:intro}

When the electric potential between the interior and exterior of a neurone exceed a certain threshold, the neurone fires. After  firing, the interior potential is abruptly reset to its rest value and the process starts over. How often it starts over depends on external stimuli (e.g. light and touch) and firing frequencies of neighbouring neurones. To better understand neurone firing, and ultimately how neurones work,  researchers \cite{gammaitoni1998stochastic,pakdaman2001coherence} use stochastic models to calculate how long it takes for the interior potential to pass the firing threshold for the first time.

Neurone dynamics is by far not the only case where first--passage problems arise. Such problems frequently occur in physics, chemistry, biology, ecology and economics \cite{benichou2011intermittent, redner} and is one of the reasons why first--passage problems are so heavily studied. But despite enormous interest there are surprisingly few cases where we know the probability distribution of first--passage times analytically.  Most cases are for Markov processes.

First--passage time densities for Markov processes mainly comes from two  approaches. In the first approach, the so--called method of images, one solves the Fokker--Planck equation with absorbing boundaries \cite{metzler2014first,risken1984fokker}.  Even though conceptually simple, it is limited to symmetric problems such as when the absorbing boundary is at the bottom of a symmetric potential well. The second approach is renewal theory \cite{vankampen,metzler2014first}. It works for non--symmetric problems but lead often to expressions in Laplace--space that cannot be inverted analytically. Even though useful, both these approaches are in practice limited to simple problems. In fact, neither of them can provide the  first--passage time density for a Brownian particle in a harmonic potential for a general boundary and starting point. Thus, in order to address more complex first--passage problems we need better analytical methods. 

Another class of useful methods have been developed to solve persistence problems. In persistence problems one wishes to know the probability $S(t)$ that a stochastic variable remains below or above a boundary from the start up to some time point $t$. The first-passage time density $\rho(t)$ is simply related to the persistence according to $\rho(t)=-dS(t)/dt$. To calculate the persistence, some researchers \cite{Sire,majumdar2,chakrabarti2009lower,verechtchaguina2006first} use methods that enumerate all trajectories with an even number of boundary crossings and calculate the probability for each trajectory. But apart from a few special cases, these crossing probabilities cannot be calculated exactly and approximations are needed.  One approximation scheme that gained popularity is the Independent Interval Approximation (IIA) \cite{McFadden,Sire2,majumdar}, which assumes  that the length of time intervals between consecutive boundary crossings are independent. However, in its present formulation the IIA assumes that the processes has a well defined velocity which means that it cannot deal with non--smooth  processes, such as Brownian motion or discrete processes. To apply IIA to those processes these shortcoming must be remedied. 

In this paper we generalise the IIA to non--smooth processes and discrete time series. Starting with the discrete case we find a simple expression for the probability density of first--passage times to a boundary from a general starting point in $z$--transform space. We then generalise our equations to encompass the continuous case and obtain a similar expression but now in Laplace transform space. The expression is based on return probability densities to the boundary and the probability that the stochastic variable is above the boundary at some time. To show the applicability of  our results we study two examples: the Ornstein--Uhlenbeck process (i.e. Brownian motion in a harmonic potential) and the discrete time Brownian walk. But our method is much more general and can in principle be used for any Markovian process and as an approximation for non--Markovian dynamics.

%%%%%%%%%%%%%%%%%%%%%%%%%%%%%%%%%%%%%%
%
%	M E T H O D S
%
%%%%%%%%%%%%%%%%%%%%%%%%%%%%%%%%%%%%%%

\section{Methods} \label{sec:method3}

\begin{figure}
\begin{center}
	\includegraphics[width=0.72\columnwidth]{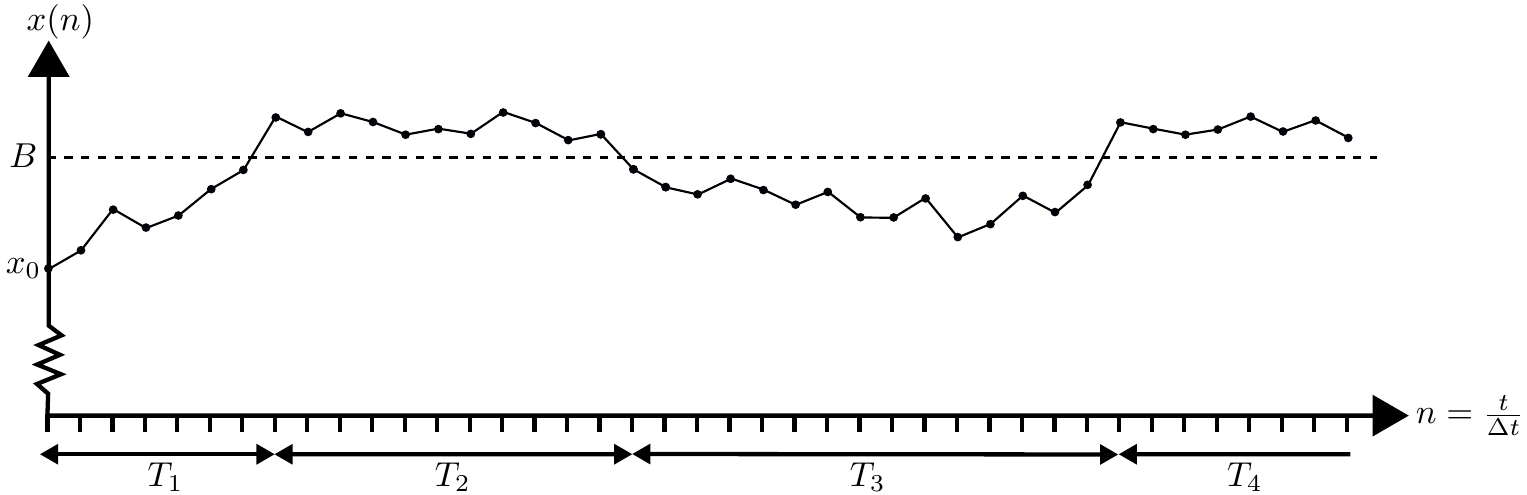}
	\includegraphics[width=0.72\columnwidth]{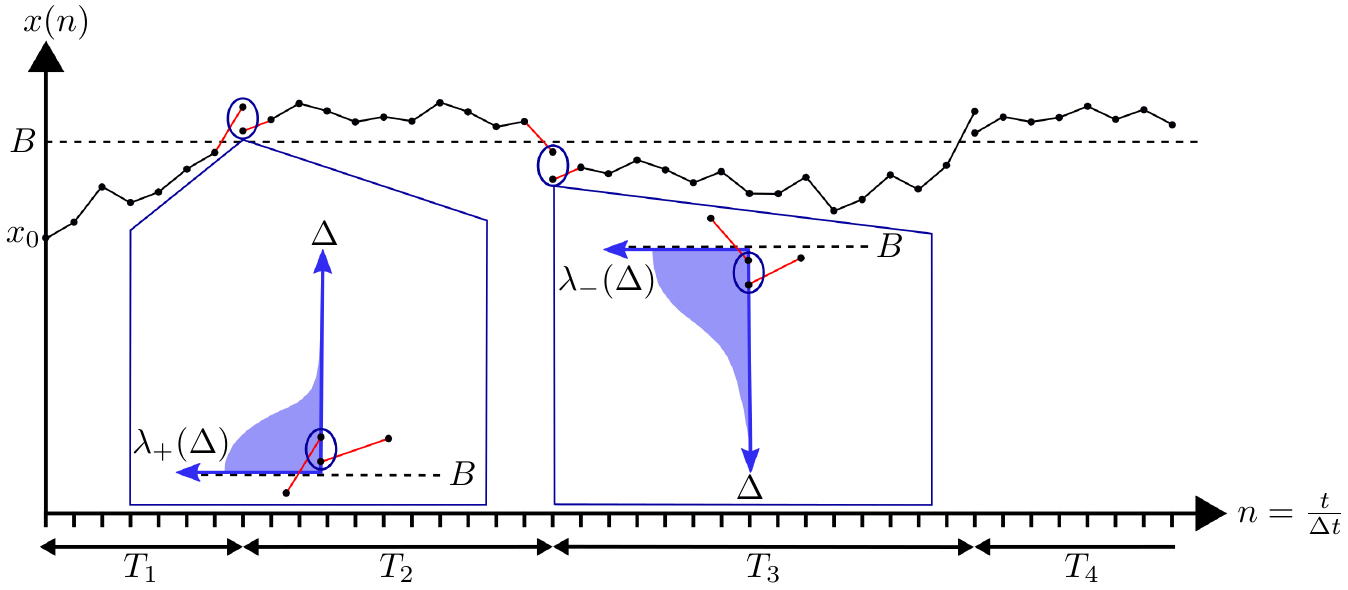}
	\caption{Discrete--time stochastic process  $x(n)$ (e.g. the position of a particle) as a function of the number of time steps $n$ ($t=n \Delta t$ where $\Delta t$ is time increment). We denote the time  spent above the boundary $B$ by  $T_2,T_4,\ldots$ and below the boundary by $T_1,T_3,\ldots$. (Top) Original process $x(n)$. (Bottom) Approximate $x(n)$  where at each crossing event we draw a new position from the overshoot distributions $\lambda_\pm(\Delta)$. Note that $\lambda_+(\Delta)$ and $\lambda_-(\Delta)$ may be different where '$+$' ('$-$') means that the process is above (below) the boundary.
	\label{fig:x(t)}}
	\end{center}
\end{figure}

In this section we outline the IIA framework and derive an expression  for the first--passage time density (FPTD) for continuous and discrete processes in one dimension. We denote the FPTD by $\fptd$ where  $t$ is time, $x_0=x(t=0)$ is the starting point of the process and $x=B$ is the location of the absorbing boundary. In discrete time we let $n$ be the number of time steps and $t=n\Delta t$ where $\Delta t$ is the time increment. To better understand the IIA approach we develop the mathematics for discrete processes and then show how it is generalised to the continuous case.

The IIA equations herein relates three core quantities. The $\fptdn$, the probability $\omega_>(n)$ that $x(n)$ is above $B$ at the $n$th time step given that $x_0<B$, and the return probability density that $x(n)$ returns to $B$ after a $B$--crossing either from above, $\overline\psi_+(n)$, or from below,  $\overline\psi_-(n)$, after $n$ steps. The quantities $\omega_>(n)$ and $\overline\psi_\pm(n)$ are inputs to our framework which one needs to calculate on case by case basis. The probability $\omega_>(n)$ is in general simple to calculate. We find it by integrating the probability density $P(x,n|x_0)$ of $x(n)$:
\begin{equation}
\omega_>(n) = \int_B^\infty P(x,n|x_0) dx.
\end{equation}
The return probability densities $\overline\psi_\pm(n)$ on the other hand are more complicated and needs to be discussed further.

To better understand $\overline\psi_\pm(n)$, consider a discrete process that pass through $B$ repeatedly (see Fig. \ref{fig:x(t)}, top). The number of steps that $x(n)$ remains below $B$ is denoted by $T_1,T_3,\ldots$, and above $B$ by $T_2,T_4,\ldots$. The number of steps to the first arrival, $T_1$, is special because it depends on $x_0$. The density of $T_1$ is  simply $\rho_B(T_1|x_0)$. After the first $B$--crossing at $T_1$, $x(T_1)$ ends up at some distance $\Delta_1\geq0$  above $B$, rarely precisely on $B$ (i.e. $\Delta_1=0)$. To calculate the distribution of $T_2$, we must consider the trajectory from $B+\Delta_1$ back across $B$. We denote the distribution of $T_2$ by $\psi_+(T_2,\Delta_1)$ where we assume that the length of $T_2$ is independent on $T_1$. This is the core assumption of the IIA and is true for Markov processes. At $T_2$, the process crossed $B$ from above and is below $B$ by $\Delta_2$. To find the number of steps until the next crossing, $T_3$, we must consider the trajectory from $B-\Delta_2$ back across $B$. The distribution of $T_3$ is $\psi_-(T_3,\Delta_2)$. Repeating this pattern we find $\psi_+(T_{2i},\Delta_{2i-1})$ ($i=1,2,3,\ldots$) for trajectories above $B$ and $\psi_-(T_{2i+1},\Delta_{2i})$ below $B$, where the $\Delta$'s are random numbers drawn from the overshoot distributions $\lambda_\pm(\Delta)$ (See Fig. \ref{fig:x(t)}, bottom).  If $\Delta$ is small with respect to $B-x_0$, overshooting the boundary by $\Delta$ will not significantly change our final results for $\fptdn$. We may therefore average $\psi_\pm(n,\Delta)$ with respect to $\lambda_\pm(\Delta)$:
\begin{equation}\label{eq:psi_bar}
 \overline \psi_-(n) = \int_0^{\infty} \psi_-(n,\Delta)\lambda_-(\Delta)d\Delta \ \ {\rm and} \ \ 
 \overline \psi_+(n) = \int_0 ^\infty  \psi_+(n,\Delta)\lambda_+(\Delta)d\Delta
\end{equation}
where the $B$ dependence enter through $\psi_\pm(n,\Delta)$. In  Appendix \ref{sec:figs}  we show simulation results for the overshoot distribution for the discrete Brownian walk which is given by
\begin{equation} \label{eq:overshoot+disc3}
\lambda(\Delta)=\sqrt{\frac{\pi}{2}}\text{erfc}\left( \frac{\Delta}{\sqrt{2}} \right).
\end{equation}

Working with the averaged return probability densities $\overline \psi_\pm (n)$ instead of $\psi_\pm(n,\Delta)$ implies that we ignore fluctuations in $\Delta$ and approximate the original process $x(n)$ by a clipped process. The dynamics of the clipped process is: when $x(n)$ crosses $B$, draw $\Delta$ from $\lambda_\pm(\Delta)$, make a  jump to $B\pm\Delta$, and continue (see Fig. \ref{fig:x(t)}, bottom). The clipped process is obviously different from the true $x(n)$ but simpler to handle analytically. But the difference is small.  We show in Appendix \ref{sec:figs} for the discrete Brownian walk that $\omega_>(n)$ for the clipped process is practically indistinguishable from the true one. Below we formulate the IIA equations based on the clipped process.

%\begin{figure}
%	\includegraphics[width=0.5\columnwidth]{one_three_cross2.pdf}
%	\caption{
%	\label{fig:clipped_process}}
%\end{figure}

Based on the clipped process we may calculate $\omega_>(N)$ ($t_N=N\Delta t$) in terms of the number of $B$--crossings. Our derivation below is the discrete time version of the derivation for the continuous time case in \cite{Sire,Sire2}. Note, however, that in \cite{Sire,Sire2} the quantity $\fptd$ does not appear, as thermal equilibrium is assumed initially. Let $p_k(N)$ be the probability for a trajectory starting in $x_0<B$ and ends up above $B$ at  $t_N$ after $k$ crossings. $\omega_>(N)$ is then  the sum of all such trajectories with odd number of crossings
\begin{equation}\label{eq:omegasum}
\omega_>(N) = \sum_{k=1,3,5,...}^\infty p_k(N).
\end{equation}
To calculate $p_1(N)$, assume that the first up--cross occurred at $n_1<N$ and that there is no down--cross between $n_1$ and $N$. Since $n_1$ can be anywhere from 0 to $N$, this gives
\begin{equation}
 p_1(N) = \sum_{n_1=0}^N \rho_B(n_1|x_0)Q_+(N-n_1)
\end{equation}
where the probability of not crossing is
\begin{equation}\label{eq:Q(t)}
 Q_\pm(n) =  1 - \sum_{n'=0}^n \overline\psi_\pm(n').
\end{equation}
To find $p_3(N)$, assume that the first up--cross occurred at $n_1$, the first down--cross occurred between $n_1$ and $n_2$,  that the second up--cross happened between $n_2$ and $n_3$, and no down--cross between $n_3$ and $N$. This gives
 \begin{equation}
 p_3(N) = \sum_{n_1=0}^N \rho_B(n_1|x_0) \sum_{n_2=0}^{N-n_1} \overline\psi_+(n_2)\sum_{n_3=0}^{N-n_2} \overline\psi_-(n_3)Q_+(N-n_3).
\end{equation}
Continuing this for $p_5(N)$, $p_7(N),\ldots$ leads to
 \begin{equation}\label{eq:p_2k-1}
 p_{2k-1}(N) = \sum_{n_1=0}^N \rho_B(n_1|x_0) \sum_{n_2=0}^{N-n_1} \overline\psi_+(n_2)\sum_{n_3=0}^{N-n_2} \overline\psi_-(n_3)\cdots
 \sum_{n_{2k-1}=0}^{N-n_{2k-2}} \overline\psi_-(n_{2k-1})Q_+(N-n_{2k-1}).
\end{equation}
Summing over all odd number of $B$--crossings we get $\omega_>(N)$ [Eq. \eqref{eq:omegasum}]. To solve Eq. \eqref{eq:omegasum} for $\fptdn$, we take the $z$--transform ($f(z)=\sum_{n=0}^{\infty}f(n)z^{-n}$), and carry out the resulting geometric series. After some algebra we obtain
\begin{equation}\label{eq:rho_z}
 \rho_B(z|x_0) = \frac{z-1}{z} \omega_>(z) g(z) \ \ {\rm where} \ \  g(z) = \frac{1-\overline{\psi}_+(z) \overline\psi_-(z)}{1-\overline{\psi}_+(z)}.
\end{equation}
%
%\begin{equation} \label{eq:g}
 %g(z) = \frac{1-\overline{\psi}_+(z) \overline\psi_-(z)}{1-\overline{\psi}_+(z)}
%\end{equation} 
%
This equation relates the first--passage time density to the probability $\omega_>(z)$ of being above the boundary, and to the return probability densities $\overline{\psi}_+(z)$ and $\overline{\psi}_-(z)$. This constitutes one of our main results in this paper. 

\subsection{Independent Interval Approximation in continuous time}

%Trajectories in continuous space that cross the boundary does it  infinitely many  times within in an infinitesimally small time interval. This is because the process is fractal. 
The proper  limit of our clipped process to a continuous time process %(no "jumps" in the trajectories) 
is the following. When  $B$ is reached from below, the trajectory makes a  jump to $B+\epsilon$, where $\epsilon$ is a small  constant. As $\epsilon\rightarrow 0$ we approach the continuous case. The overshoot distributions $\lambda_\pm(\Delta)$ for this cases is a Dirac delta function, which leads to 
\begin{equation} \label{eq:clipped+cont}
 \overline \psi_\pm(t) = \lim_{\epsilon \rightarrow 0 } \int_{-\infty} ^\infty\psi_\pm(t,\Delta)\delta\left(\Delta-(B\pm \epsilon)\right)d\Delta .
\end{equation}
In Appendix \ref{sec:figs} we show explicitly how the clipped Ornstein--Uhlenbeck process convergence to the continuous one as $\epsilon$ gets smaller.

To derive the IIA equations in the continuous case, we proceed in the same way as for the discrete case using $p_k(t)$, but with sums in Eq. \eqref{eq:p_2k-1} changed to integrals ($\sum_n\rightarrow \frac{1}{\Delta t}\int dt$) as we let $\Delta t\rightarrow 0$ and $n\rightarrow \infty$, while maintaining $t=n \Delta t$ constant. If we take the Laplace transform ($f(s)=\int_0^{\infty}f(t)e^{-st}\,dt$) of the sum over $p_k(t)$ [Eq. \eqref{eq:omegasum}] we obtain a similar geometric series as before that leads to 
\begin{equation} \label{eq:main+result}
\rho_B(s|x_0) = s  \omega_>(s)  g(s) \ \ {\rm where} \ \ g(s) = \frac{1-\overline \psi_+(s)\overline \psi_-(s)}{1-\overline \psi_+(s)}.
\end{equation}

\section{Results}
In this section we apply our main results, Eqs. \eqref{eq:rho_z} and \eqref{eq:main+result}, to two examples. The first example is the continuous Ornstein--Uhlenbeck process, and the second one is the discrete Brownian walk. We also show that our results lead to Kramers escape for a general Gaussian stationary process (Appendix \ref{sec:kramers}), and that our method is consistent with the method of images (Appendix \ref{sec:MMI}). To test the validity of our theoretical results we compare them to Langevin dynamics simulations (see Appendix \ref{sec:sim} for simulation details). 

\subsection{Application 1: Ornstein--Uhlenbeck process} \label{se:cont+time}

The FPTD for the Ornstein--Uhlenbeck process (OUP) is inherently difficult to calculate explicitly \cite{grebenkov2015first}. The one exception is the symmetric case when the absorbing boundary is at the bottom of the harmonic well which can be solved with the method of images \cite{blake}. However, this  does not work for the general problem. Instead, several efforts focused on the renewal equation in Laplace space. But because the renewal equation cannot generally be inverted analytically \cite{siegert,darling} researchers used numerical inversion \cite{mullowney} and series expansion around poles \cite{alili}. The last example \cite{alili} currently holds the best analytical approximations for the FPTD in the field. But even though in principle exact, none of their expressions are on closed form and must be evaluated numerically. To work with those expressions one must specify at least one cut--off parameter (sometimes two) which in practise must be done by trail and error. We compare our results to \cite{alili} using their so--called integral representation (see appendix \ref{sec:alili} for explicit details). Our approach does not share the cut--off parameter problem because we start in another end. We rely on a particular functional form of $\overline \psi_+(t)$ that is asymptotically true for Markovian Gaussian Stationary Processes (GSPs); They  decay exponentially. The return probability densities, together with the probability density $P(x,t|x_0)$, that we know, yields our final expression. %Below we derive this expression starting by discussing the return probability densities.

\subsubsection{Analytical predictions}
To calculate the return probability densities, imagine first a symmetric GSP where $x(t)$ behaves in the same way below and above the boundary $B=0$. For Markovian GSPs we know the probability that $x(t)$ does not change sign during the time interval $[0,t]$ (i.e. the persistence) given that the process started in the infinite past such that it is stationary at $t=0$. It is given by \cite{majumdar2}
\begin{equation} \label{eq:doob+theorem}
Q(t) = \frac{2}{\pi}\,\text{arcsin}\left(  e^{ -rt  } \right)\simeq  e^{ -rt  }. %\quad0\leq t < \infty
\end{equation}
Because of symmetry, the return probability densities $\overline  \psi_\pm(t)$ are in this case equal. Denoting them by $\overline  \psi(t)$ and using that $\overline \psi(t) = - dQ(t)/dt$  (because $Q(t) = 1 -\int_0^t\overline \psi(t')dt'$, see Eq. \eqref{eq:Q(t)}), we obtain that $\overline \psi(t)$ decays exponentially% with rate one (dimensionless variables)
\begin{equation}
\overline  \psi(t)  \simeq re^{- rt}.
\end{equation}
To generalise this expression to the asymmetric case $B\neq 0$ where $\overline \psi_{-}(t)\neq\overline \psi_{+}(t)$, we follow  \cite{Sire,slepian} and introduce crossing rates $r_\pm$ from above and below $B$ 
\footnote{Note that there is no dependence of $x_0$ on $\overline \psi_{\pm}(t)$. They depend only on the location of the boundary $B$ and thus on the crossing rates $r_{\pm}$. Therefore we can work with a general $x_0$.}. This leads to
%
%Thus, the probability of not crossing the boundary from above must be different from below. Therefore we define the  two persistence functions with respect to $B$ in their asymptotic limits
%
\begin{equation} \label{eq:persistence+B}
Q_{\pm}(t) \simeq \text{exp}\left( -r_{\pm}t  \right) 
\end{equation}
and
\begin{equation} \label{eq:psi+B}
\overline  \psi_{\pm}(t) \simeq \,r_{\pm}\text{exp}\left( -r_{\pm}t  \right). 
\end{equation}

With $\overline  \psi_{\pm}(t)$ at hand, we may calculate the FPTD  for a general Markovian GSP from  Eq. \eqref{eq:main+result}. First, we use that $\overline \psi_\pm(s) = 1/(1+s/r_\pm)$ in Laplace space. Second we put $\overline \psi_\pm(s)$ in $g(s)$ [Eq.  \eqref{eq:main+result}] so that $ g(s)=1+r_+/(r_-+s)$. After inversion we find
%
%\begin{equation} \label{eq:g+invert}
%g(t) = \delta(t) + r_+e^{-r_- t}
%\end{equation}
%
%Second, using this we invert which gives the FPTD
%
\begin{equation} \label{eq:oup4}
\fptd = \frac{d \omega_>(t)}{dt} + r_+\int_0^t  e^{-r_-t'}\frac{\partial\omega_>(t-t')}{\partial t} \,dt'.
\end{equation}
This equation is valid for any Markovian GSP. To get further, we need to address specific examples where we can calculate $r_\pm$ and $\omega_>(t)$.

For the OUP, $\omega_>(t)$ is given by
\begin{equation} \label{eq:omega+oup}
\omega_>(t)=\frac{1}{2}\text{erfc}\left( \frac{B-x_0e^{-t}}{\sqrt{2(1-e^{-2t})}} \right).
\end{equation}
It is calculated from $\int_B^\infty P(x,t|x_0) dx$ where \cite{vankampen} (dimensionless variables, see Appendix \ref{sec:sim})
\begin{equation} \label{eq:oup2}
P(x,t|x_0)= \frac 1  {\sqrt{2\pi (1-e^{-2t})}} \exp \left( -\frac{(x-x_0e^{-t})^2}{2(1-e^{-2t})}  \right).
\end{equation}
To calculate  crossing rates $r_\pm$, we proceed as follows. First, we get $r_+$ from the normalisation condition $\int_{-\infty}^{\infty}\fptd\,dt=1$,  or ${\rho}_B(s\to0|x_0)=1$. Also, using ${g}(s\to 0) =1+r_+/r_-$ and $\lim_{s\to0}s\omega_>(s)=\omega_>(t\to\infty)$ in Eq. \eqref{eq:main+result} leads to
\begin{equation} \label{eq:tau+plus+norm}
%r_+ = r_-\left(\frac {\int_{-\infty}^{\infty} e^{-V(x)}\,dx} {\int_B^{\infty} e^{-V(x)}\,dx} -1\right)
%r_+ = r_-\left( \frac{1}{\omega_>(\infty)}-1  \right)
r_+ = r_- \left( \frac 2 {\text{erfc}\left( B/\sqrt 2\right)} -1\right).
%r_+ = r_-\left( \frac{2}{\text{erfc}\left( \frac{B-\mu(x_0,t\to\infty)}{\sqrt{2\sigma^2(t\to\infty)}} \right)}-1  \right)
\end{equation}

The other crossing rate $r_-$ is equal to the inverse of the mean--first passage time $\tau$ to $B$ from $x_0=0$, where the latter result follows from the backward Fokker--Planck equation \cite{gardiner}
\begin{equation} \label{eq:mfpt+bfpe+ou}
r_- = \frac 1 \tau, \ \ \ {\rm where } \ \ \ \tau =  \int_0^B dz \,e^{z^2/2}\int_{-\infty}^zdy\,e^{-y^2/2}.
\end{equation}
This can be understood as follows. When $B$ is far away from the potential minimum $x=0$, up--crossing events are rare. When they are rare, the distribution of times between up--crossing events is approximately the same as the FPTD. Because the process's equilibration time is much shorter than the mean--first passage time to $B$, $x_0$ is approximately equilibrated with average $x_0=0$
\footnote{The equilibrium time is of order one whereas $\tau$ is about ten times larger already when $B\approx 2$ where $\tau \sim \exp(B^2/2)$.}.
This  means that the FPDT is asymptotically Kramers expression  $\rho_B(t|0) \simeq  \tau^{-1}e^{-t/\tau}$ \cite{hanggi}. This implies that $\overline\psi(t)\simeq \rho_B(t|0)$ with $r_-=1/\tau$.

Before we compare Eq. \eqref{eq:oup4} with simulations we clarify some of its limitations. First, note that Eq. \eqref{eq:main+result} is exact within the IIA but in general we do not know $\overline\psi_\pm(t)$. If we assume that  $\overline\psi_\pm(t)$ decays exponentially we arrive at Eq. \eqref{eq:oup4}. This assumption is only asymptotically true. This means that Eq. \eqref{eq:oup4} is less accurate when $B$ and $x_0$ are too close. If they are, several crossing events occur  at short times and $\overline \psi_\pm(t)$ is not decaying as a single exponential. To better understand what is meant by 'short times' we see in Fig. \ref{fig:one_B} that $B=3$ and $x_0\leq 0$ gives good agreement with Langevin dynamics simulations for $\fptd$, while there is a discrepancy for short times when $B=3$ and $x_0\geq 1$. To improve the results for short times we could in principle take into account  sub--leading terms $\overline \psi_\pm(t) = r_{\pm}e^{ -r_{\pm}t } [1 + a_{\pm,1} e^{-(r_{\pm,1}-r_\pm)t}+a_{\pm,2} e^{-(r_{\pm,2}-r_\pm)t}+...]$. However, this would introduce new parameters as well as conditions for when to truncate the sum. Using the asymptotic behaviour of $\overline \psi_\pm(t)$ makes our approach free of cut--off parameters.

%Second, we point out that the survival probability $Q(t)$ in Eq. \eqref{eq:doob+theorem} in principle depends on $x_0$. We omit this correction and use for simplicity the asymptotic form of $\overline \psi_\pm(t)$ that is independent of $x_0$.

\subsubsection{Simulations and numerical results} \label{sec:harmonic}

\begin{figure}[] 
	\includegraphics[width=0.32\columnwidth]{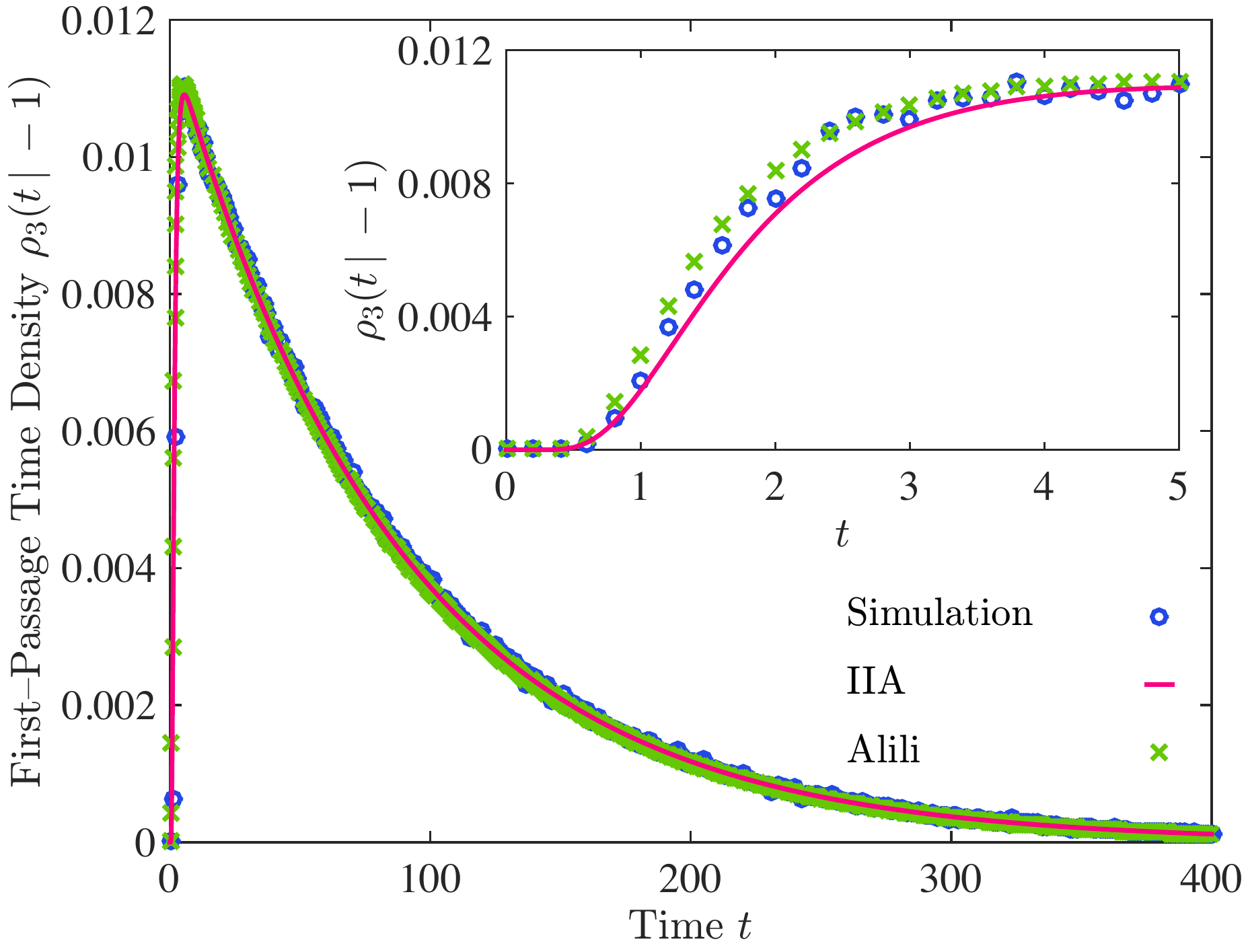}
	\includegraphics[width=0.32\columnwidth]{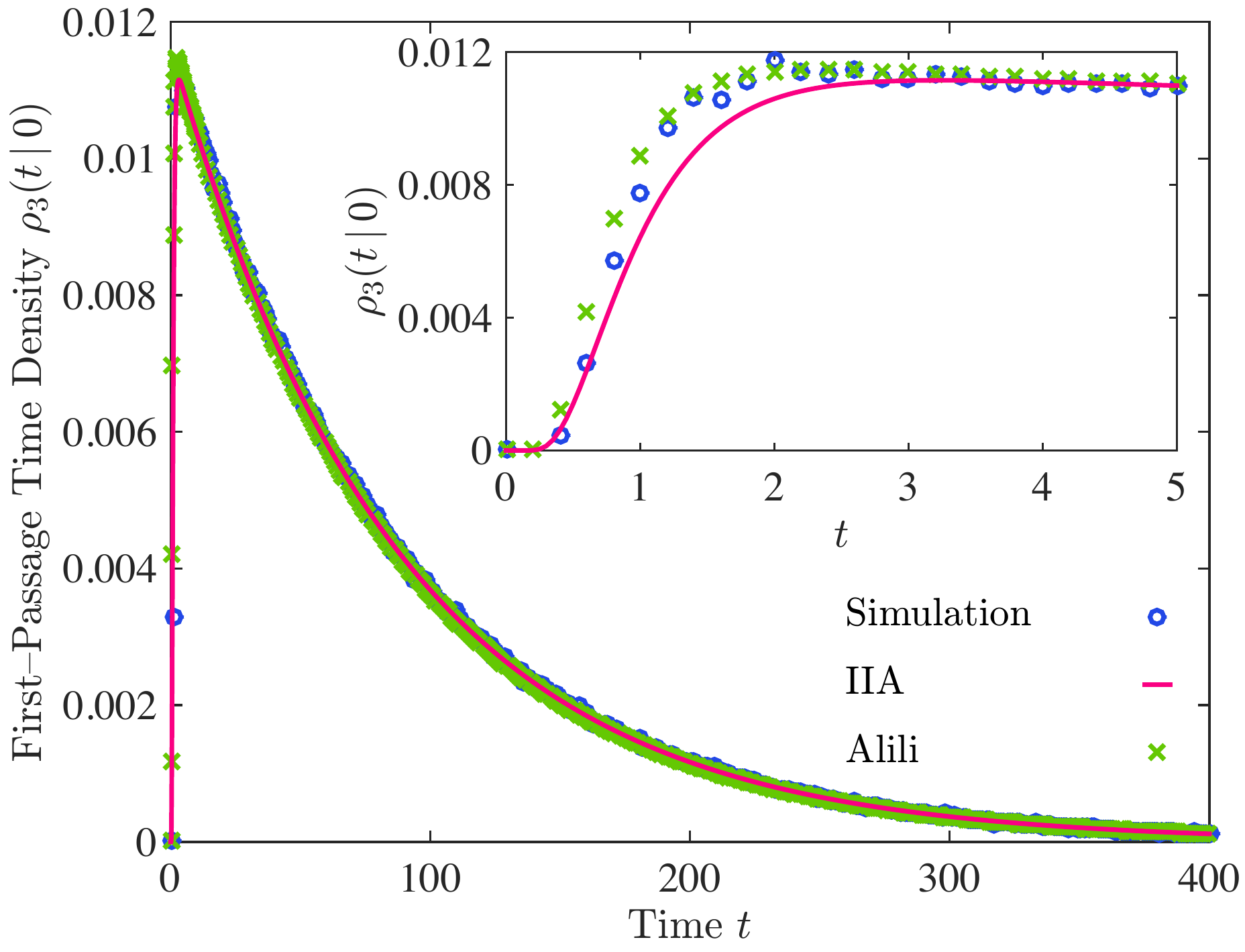}
	\includegraphics[width=0.32\columnwidth]{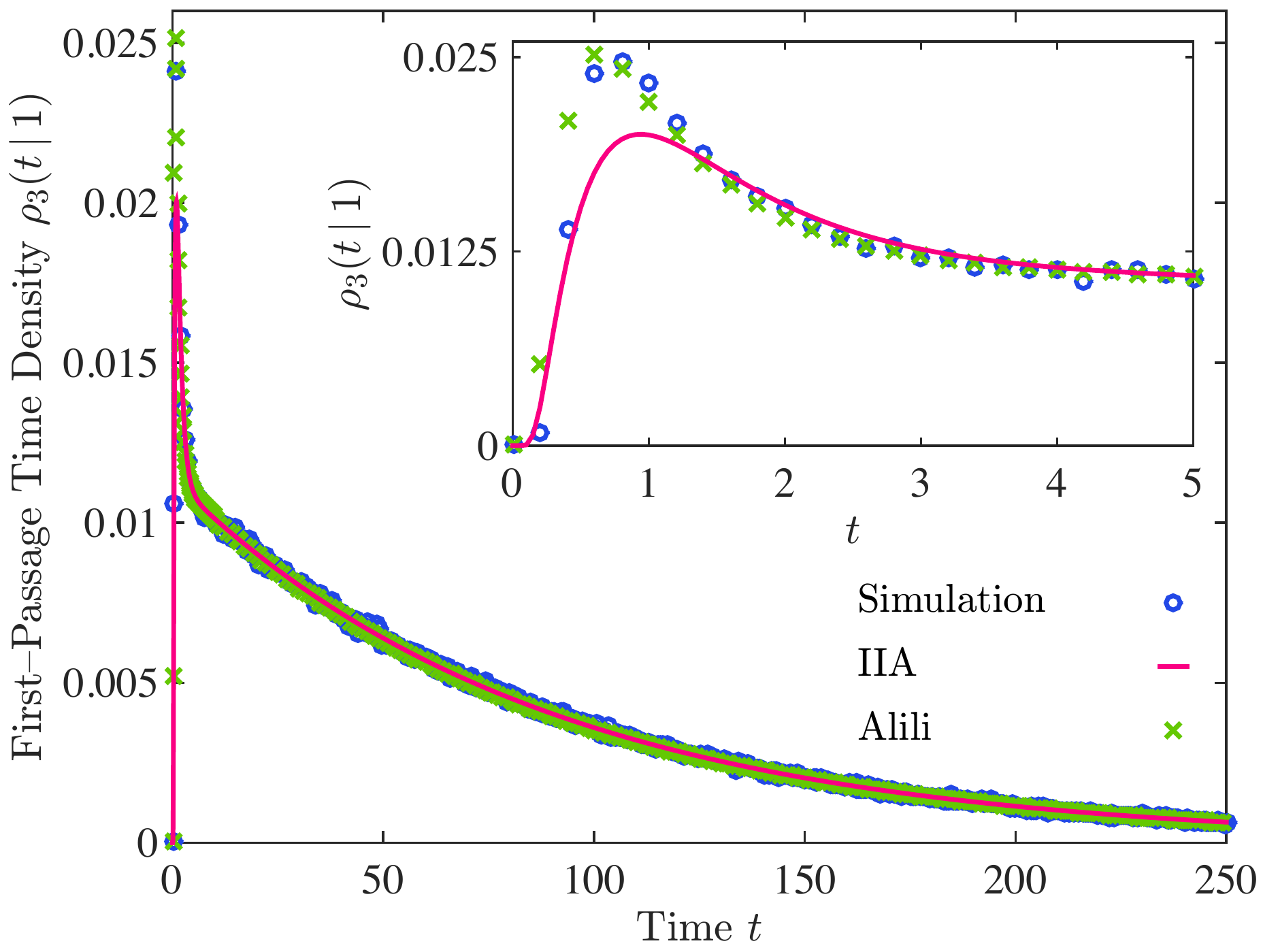}
	\caption{First--passage time density $\fptd$ out of a harmonic well $V(x)=x^2/2$ when the boundary is at $B=3$ and $x_0=-1$ (left), $x_0=0$ 	(middle), $x_0=+1$ (right).  The solid line is  Eq. \eqref{eq:oup4}, 'o' are results from Langevin dynamics simulations (averaged over $10^6$ realisations), and '$\times$' is the approximation in \cite{{alili}} [Eq. \eqref{eq:alili+approx}]. The insets show the behaviour at short times.% (the bin size is five times smaller then in the main panel). 
	}
\label{fig:one_B}
\end{figure}

To validate our method we compare $\rho_B(t|x_0)$ to Langevin dynamics simulations. 
%(the simulation scheme is explained in Appendix \ref{sec:sim}). 
In Fig. \ref{fig:one_B} we show $\rho_B(t|x_0)$ for different $x_0$ keeping $B=3$ fixed. The simulation results are represented by circles and the IIA Eq.  \eqref{eq:oup4}  by solid lines. The main panels show the overall behaviour whereas the insets show the short time dynamics. Overall, we get good agreement with simulations. 

In the long time limit, our analytical FPDT decays exponentially. This agrees with Kramers escape rate, and our IIA equations capture this regime well for all values of $B-x_0$. For short times, there is a discrepancy between simulations and our IIA equations when $B-x_0$ gets smaller. This is because  a considerable amount of first--passage events occur for short times before $\overline \psi_\pm(t)$ has  attained its single exponential form, as discussed above. Also, the reason why our IIA equations systematically underestimates the simulations for short times is the following. Assume for simplicity that$x_0=0$ and that $B$ infinitesimally above $x_0$. In this case, the first--passage dynamics is practically indistinguishable from Brownian motion for short times. For Brownian motion we can show that $\overline\psi_\pm(t)$ is a Dirac delta function (Appendix \ref{sec:MMI}) which means that there is an infinite number of boundary crossings once the boundary is crossed. This is very different from $\overline\psi_\pm(t)\simeq e^{-r_\pm t}$ that says that the average time between two boundary crossings is $1/r_\pm$.

In Fig. \ref{fig:one_B} we also included one of the best  analytical approximations  for the FPTD \cite{alili} (see  Appendix \ref{sec:alili}). Their formula approximates the short time dynamics better than our method while for long times both approaches match well with each other. %Although successful, Alili's result is tedious compared to our result.

\subsubsection{First--passage through two boundaries}
\begin{figure}[]
	\includegraphics[width=0.32\columnwidth]{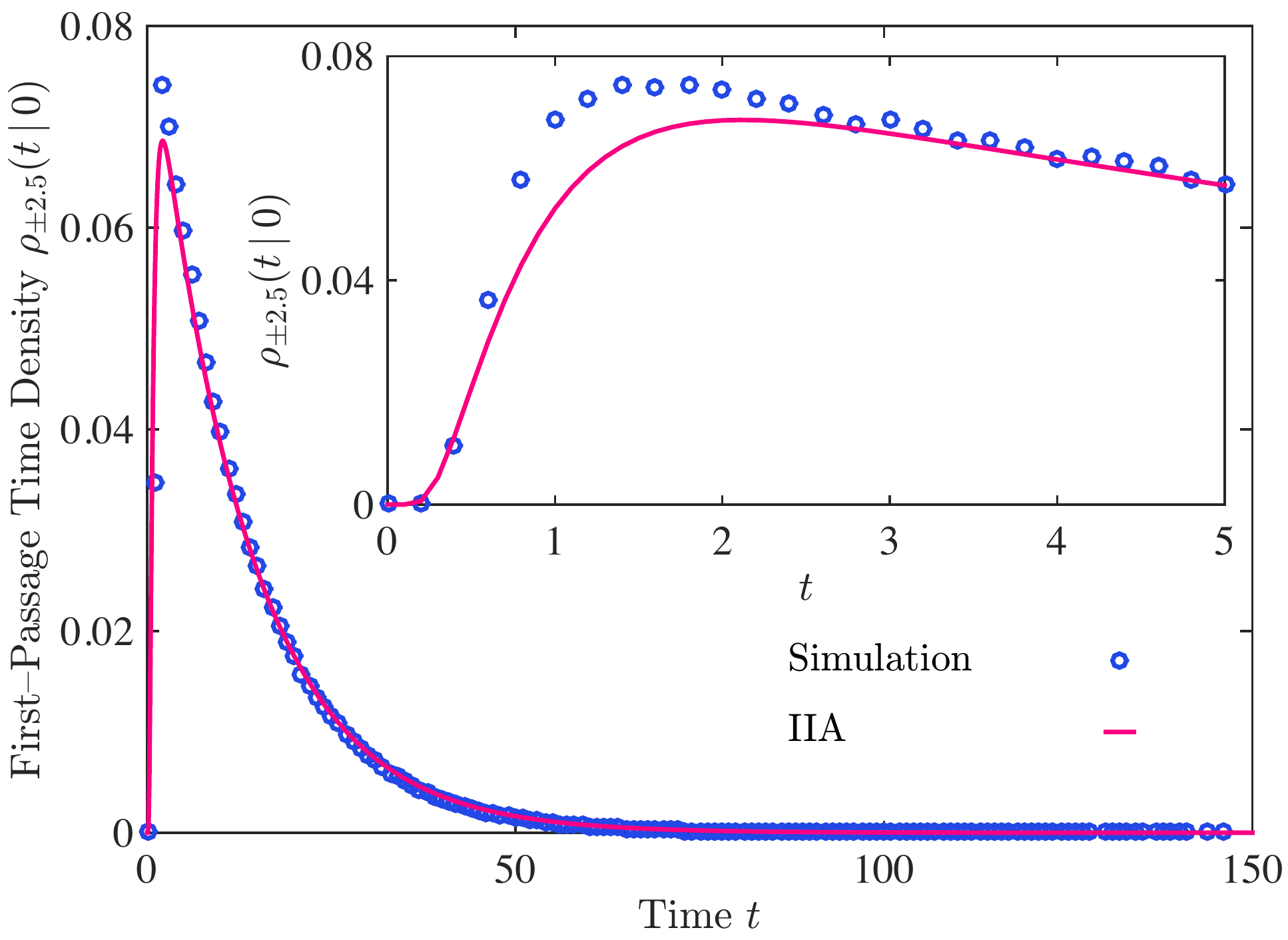}
	\includegraphics[width=0.32\columnwidth]{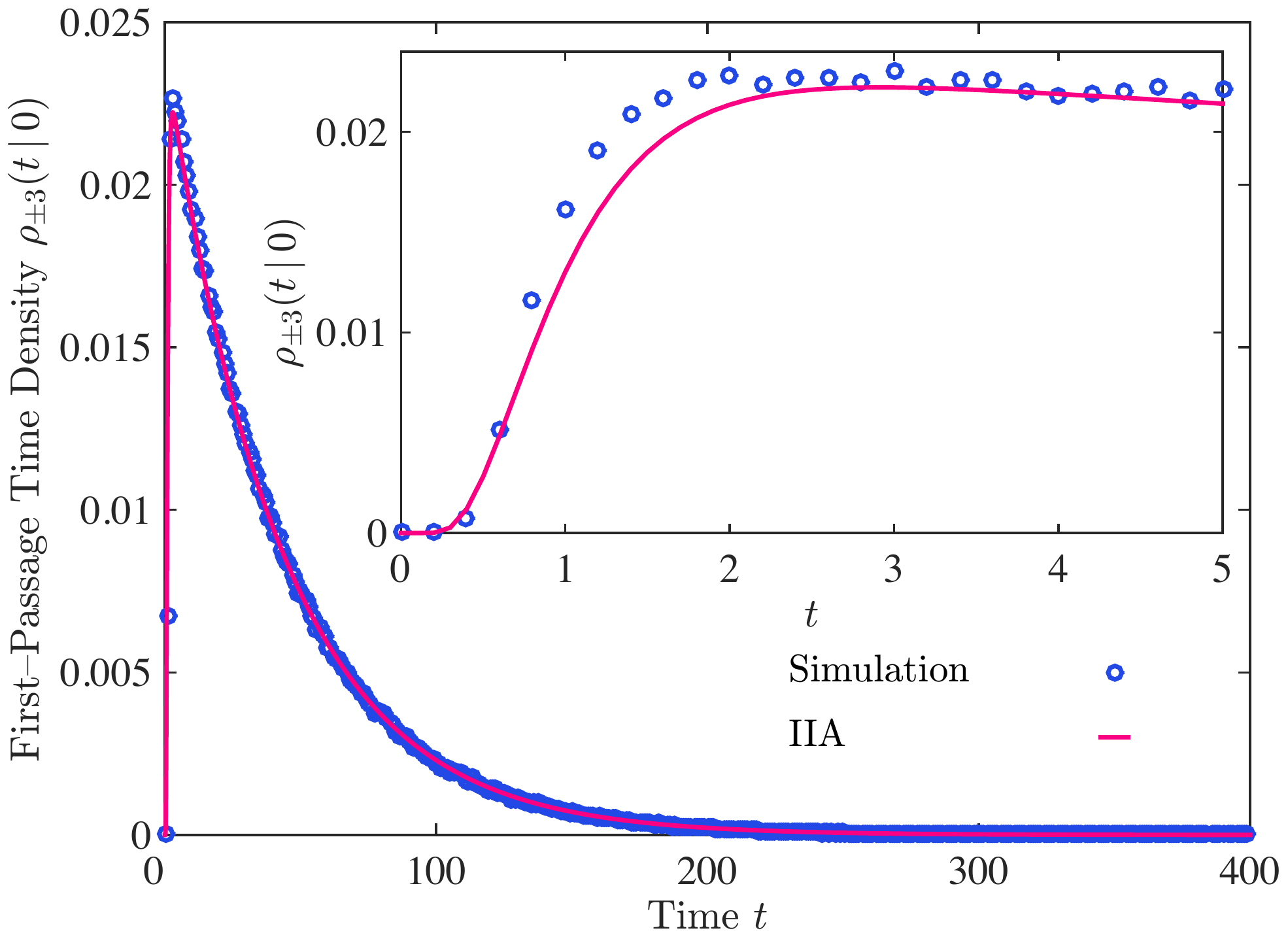}
	\includegraphics[width=0.32\columnwidth]{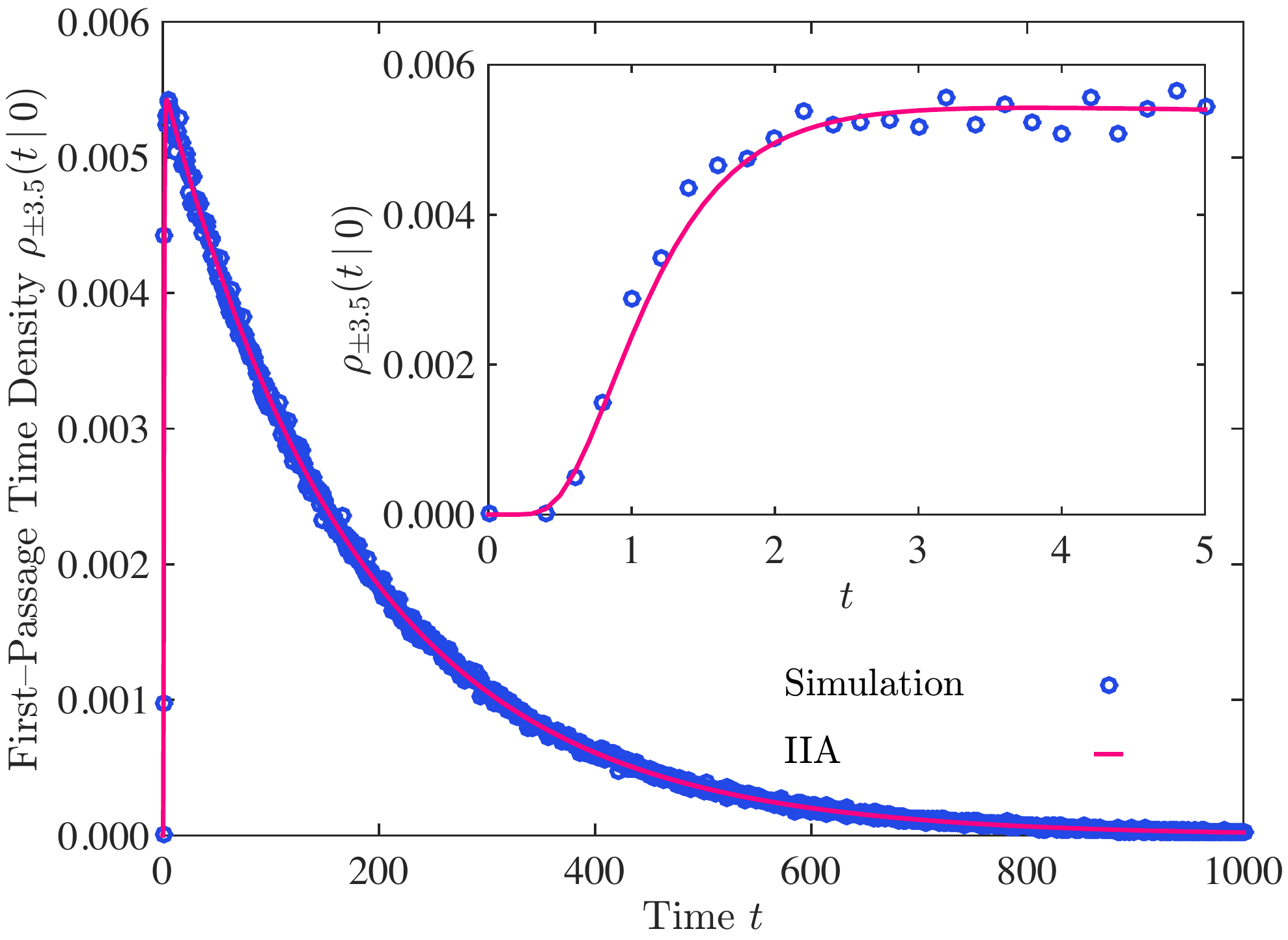}	
	\caption{First--passage time density $\fptd$ out of a harmonic well $V(x)=x^2/2$ with two boundaries and $x_0=0$: 	(left) $B=\pm  2.5$, (middle) 	$B=\pm 3$, (right) $B=\pm 3.5$. The solid line is  Eq. \eqref{eq:oup4},  and 'o' are 	results from Langevin dynamics simulations (averaged over	$10^6$ 	realisations). The insets show the 		behaviour at short times.}
	\label{fig:two_B}
\end{figure}

It is straightforward to generalise our method to two boundaries, $B>0$ and $B'<0$. Clearly, if $B\neq |B'|$  we need four  return probability densities which our method is unable to handle. But for the symmetric case, $B=|B'|$,  $\overline\psi_\pm(t)$ are enough to describe the crossing in and out of the region $-B\leq x\leq B$. To calculate $r_-$ we use a generalisation of Eq. \eqref{eq:mfpt+bfpe+ou} to two boundaries (see e.g. \cite{gardiner}), and to get $r_+$ we use Eq. \eqref{eq:tau+plus+norm} where we replace $\omega_>(\infty)\rightarrow 2\omega_>(\infty)$ (the probability that $\left| x(t) \right| > B$). Figure \ref{fig:two_B} shows that our IIA equations match simulations for two boundaries with similar accuracy as for one boundary.

\subsection{Application 2: Discrete time Brownian walk} \label{sec:disc+time}
The Brownian walk \cite{majumdar2} is one of the simplest cases of a non--stationary process. Nevertheless, its FPTD is not known for general $B$ and $x_0$ except in terms of a double Laplace transform \cite{majumdar2} that no one thus far have been able to invert. The one exact result that exists is the  Sparre Andersen theorem \cite{sparre} that says that the persistence to stay above (or below) the boundary when $B=x_0$ is
\begin{equation} \label{eq:sparre_thm}
q(n) = \binom{2n}{n}2^{-2n}.
\end{equation}
Here we use $q(n)$ and the IIA formalism to put forward a simple summation formula for the FPTD for general $B$ and $x_0$. As before, we first need to find the return probability densities $\overline\psi_\pm(n)$.

\subsubsection{Analytical predictions}
In this problem $x(n)$ behaves in the same way on both sides of the boundary, which means that the return probability densities on either side of $B$ are equal, $\overline\psi_{\pm}(n) = \overline \psi(n)$. We approximate $\overline\psi(n)$ with the discrete derivative of $q(n)$, that is $\overline \psi(n)\approx-\left[q(n) - q(n-1)\right] \Theta(n-1)$. Here  $\Theta(n)$ is the unit step function (discrete heavy side step function) that takes care of the initial condition $\overline \psi(n=0)=0$. From Eq. \eqref{eq:sparre_thm} it follows that
\begin{equation} \label{eq:AR(1)81}
\overline \psi(n) \approx% \left[q(n-1) - q(n)\right] \Theta(n-1) =
 %\frac{\Gamma(n-1/2)\Theta(n-1)}{2\sqrt{\pi}\,\Gamma(n+1)}=\frac{2^{-n}}{2n-1}\frac{(2n-1)!!}{n!}
 \frac{2^{n}}{2n-1}\frac{n!}{(2n)!!}q(n) \Theta(n-1).
\end{equation}
If we put $\overline \psi_\pm(n)=\overline \psi(n)$, Eq. \eqref{eq:rho_z} yields $g(z)=1+\overline \psi(z)$. Using this in $\rho_B(z|x_0)$ [Eq. \eqref{eq:rho_z}] and inverting to $n$--space leads to 
\begin{equation} \label{eq:rho_n1}
\rho_B(n|x_0) = \omega_>(n) + \sum_{k=1}^{n-1}\omega_>(n-k)\left[\overline \psi(k)-\overline \psi(k-1)-\delta_{k,1} \right],
\end{equation}
where $\omega_>(n)=\frac 1 2\text{erfc}\left[(B-x_0)/(\sqrt{2n}) \right]$ is calculated in the same way as Eq. \eqref{eq:omega+oup}. Equation \eqref{eq:rho_n1}  is a generalisation of the Sparre--Andersen theorem to general boundary and initial conditions.

For long times we expect that $\rho_B(n|x_0)\simeq \overline \psi(n)$. Indeed,  expanding Eq. \eqref{eq:AR(1)81} for large $n$ we get the Brownian walk result  $\rho_B(n|x_0)\sim n^{-3/2}$.

\subsubsection{Simulations and numerical results}

\begin{figure}[] 
	\includegraphics[width=0.32\columnwidth]{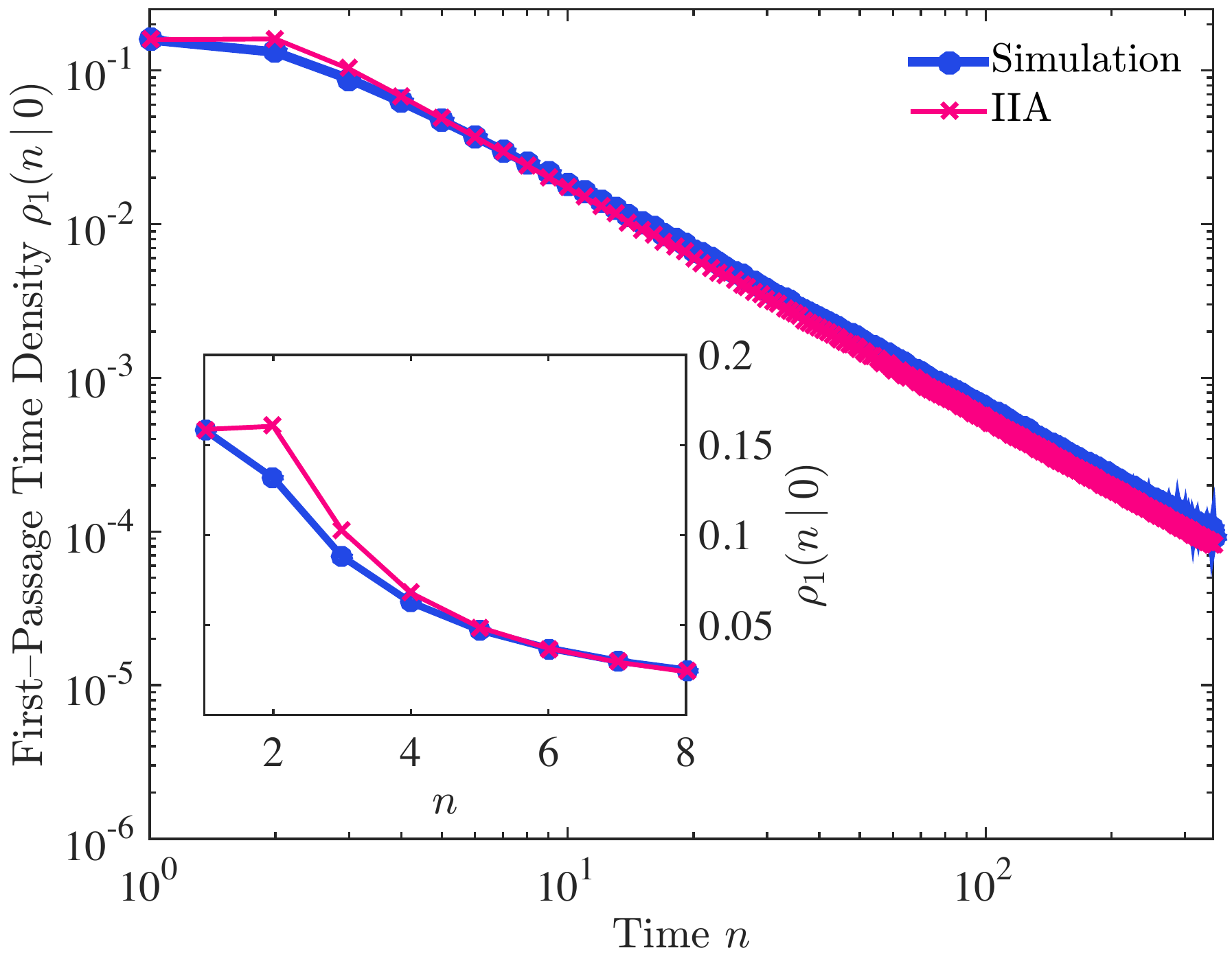}
	\includegraphics[width=0.32\columnwidth]{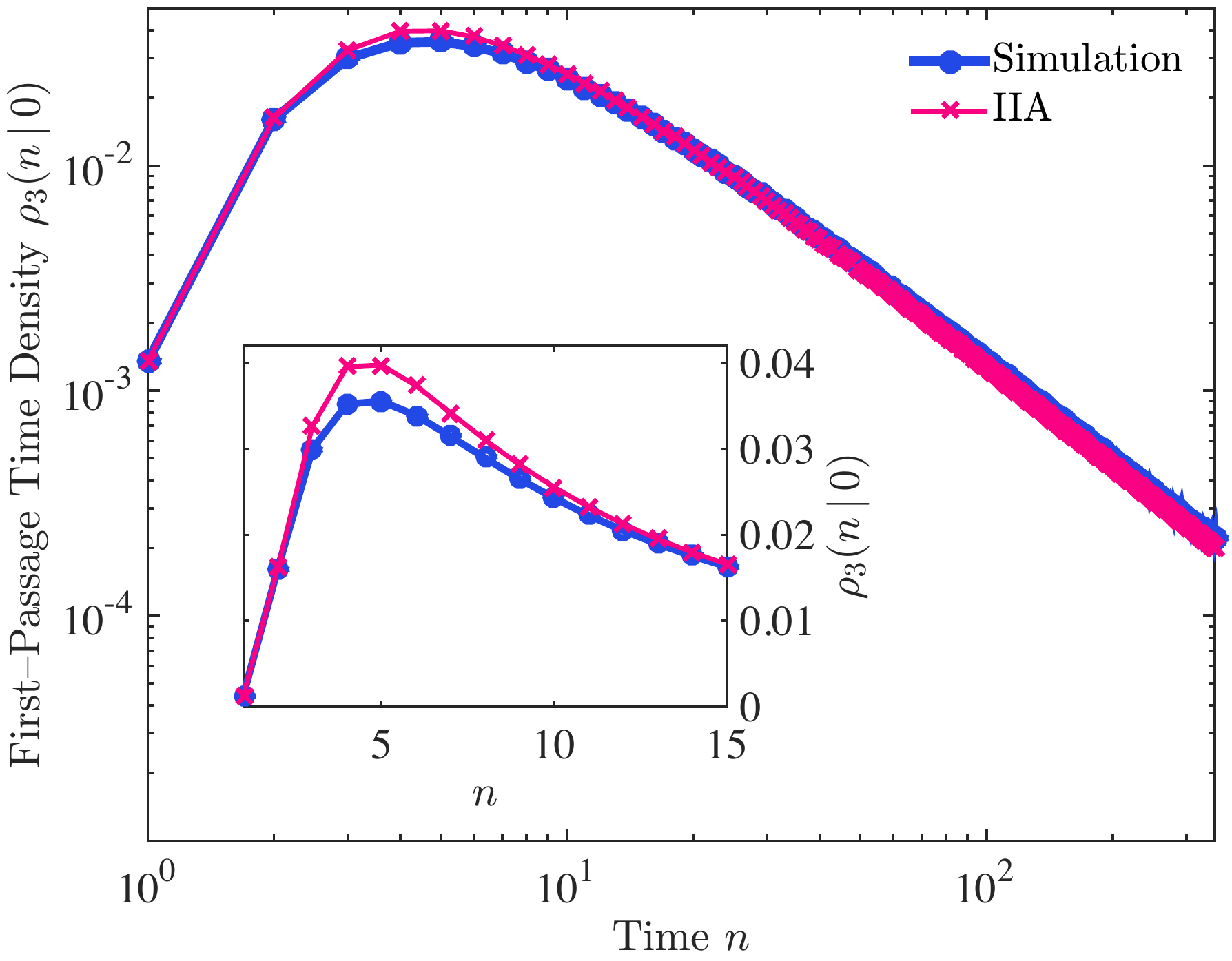}
	\includegraphics[width=0.32\columnwidth]{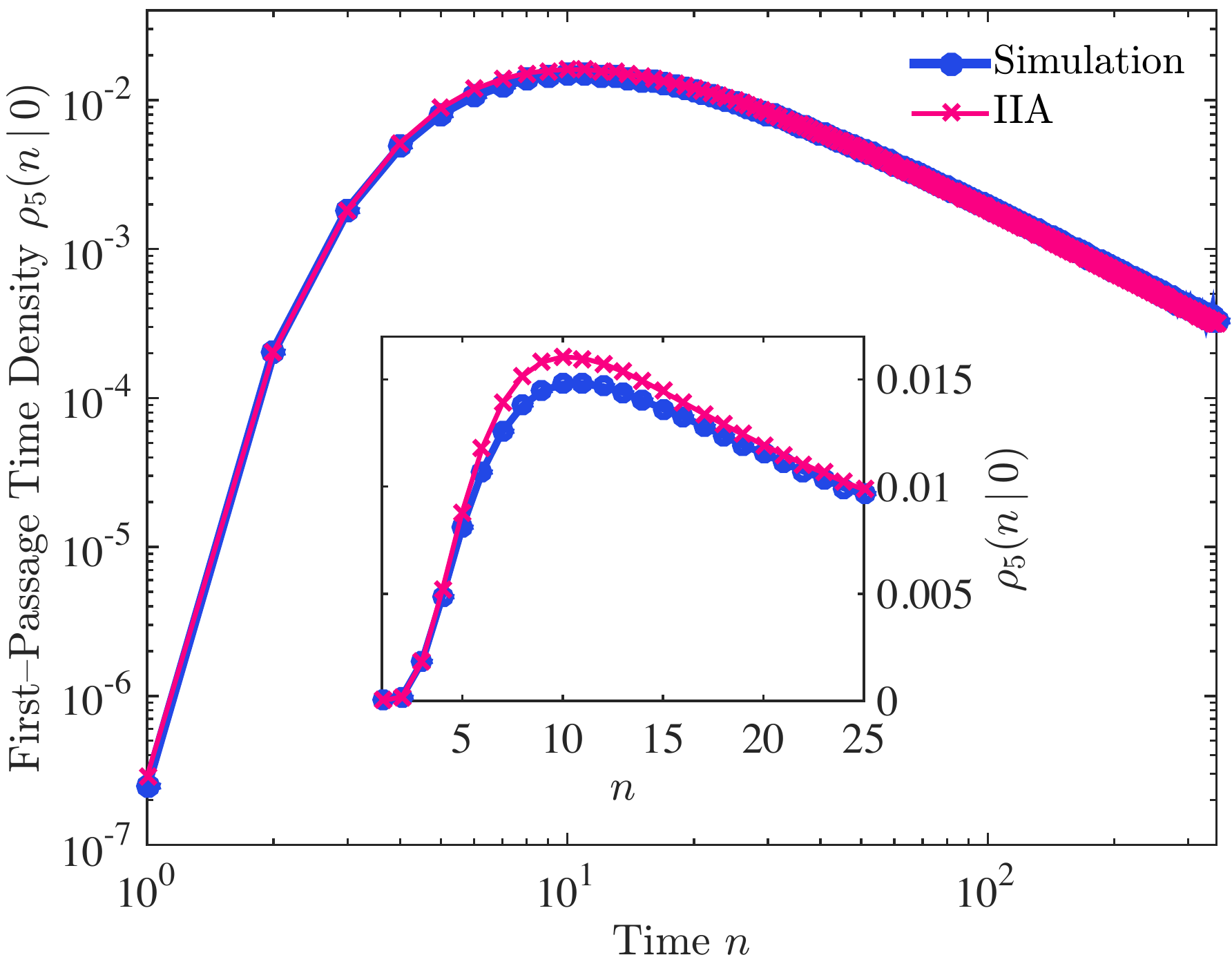}
	\caption{First--passage time density $\rho_B(n|x_0=0)$ for the simple Brownian walk when the boundary is at: $B=1$ (left), $B=3$ (middle), $B=5$ (right). Connected crosses come from Eq. \eqref{eq:rho_n1} while connected rings represent simulations (averaged over $2\times10^7$ realisations). Insets display  short--time dynamics. \label{fig:one_B2}}
\end{figure}

In Fig. \ref{fig:one_B2} we compare Eq. \eqref{eq:rho_n1} to simulations. Overall we find good correspondence, especially as $B-x_0$ increases. But as it decreases, we start to see deviations for small times,  e.g for $B-x_0=1$. The reason is that the overshooting  start to play a role and the derivative of the persistence is no longer a good approximation to $\overline\psi(n)$. From simulations we find that the average overshooting length is 0.626... that is comparable to $B-x_0=1$.

%%%%%%%%%%%%%%%%%%%%%%%%%%%%%%%%%%%%%%
%
%	DISCUSSION & CONCLUSION
%
%%%%%%%%%%%%%%%%%%%%%%%%%%%%%%%%%%%%%%

\section{Summary  and outlook}
There are  plenty of examples where one wants to know the probability density of first--passage times to a boundary. To find this density, one often use the method of images or renewal theory. These approaches are however in practise limited to simple cases. To find better methods, we improved the so--called Independent Interval Approximation (IIA), developed for survival probability problems, that is limited to smooth stochastic processes where trajectories have well defined velocities. This excludes for example Brownian motion. We generalised IIA to continuous and discrete non--smooth trajectories. From our IIA formalism we  derive a simple expression for the first--passage time density to a general boundary and initial condition in one dimension. This expression relies on that we know the functional form of the return probability densities. But once it is known our approach is parameter free. To show the validity of our expression we applied it to the Ornstein--Uhlenbeck process (OUP) and the discrete time Brownian walk. For the OUP we use that the return probability densities decays exponentially for long times \cite{majumdar2}, $\sim \text{exp}\left(-r_{\pm}t\right)$, where we identify $r_-$ (up--crossing rate) as the reciprocal of the mean first--passage time, which is known \cite{gardiner}. Then, from the normalisation condition of the first--passage time density we determine the last parameter, $r_+$ (down--crossing rate). In discrete time we apply our IIA formula to the Brownian walk where we use the Sparre--Andersen theorem which yields the return probability densities. Both cases match well with Langevin dynamics simulations. We also show that (i) our approach reproduces Kramers expression for escape of a Brownian particle out of a harmonic potential, and (ii) that it becomes equivalent to the method of images for symmetric problems, e.g. when the boundary is at the bottom of a potential well.

Our IIA equations are new, and we anticipate that they will have a wide applicability to previously intractable  first--passage and escape problems. %In particular, it would be interesting to see how our method performs for generalised Gaussian processes, e.g fractional Brownian motion. At first glance this does not look promising because time intervals between crossings are correlated which violates the very core of IIA. However, simple scaling arguments that relate the number of boundary crossings to the first--passage time density yields correct exponents.

%In the simulations of the FPTD, the initial position has been chosen to $x_0 = 0$. The reason for this is that for the harmonic case, the Brownian particle will reach its stationary state, with mean position zero, exponentially fast, see Eq. \eqref{eq:harmonic+pdf}. Thus, having the initial position $x_0\neq 0$ will most likely render in crossing the origin before crossing $x = B$. %Moreover, the equations simplify when $x_0=0$. 

%In the simulations we have used a time step $\Delta t=0.005$ and averaging has been done over $10^7$ realizations. The exact MFPT is known by use of the BFPE. However, the generated data overestimates this value and one need a time step $\Delta t\to0$ in order to correct this. Therefore the value given to the constant parameter $r_-=1/\tau$ is taken from the simulations. But this has no affect on the accuracy of our integral formula or on the result as a whole. There are, however, alternative simulation methods to use where one searches for the first--passage on a binary tree using a bridge between two known space-time coordinates which seems to give much better results \cite{recursive}. 

%our model when the OUP is a Markov process. Since time intervals for a Markov process are uncorrelated, the IIA becomes exact. However, due to the lack of "nice" Laplace transformable functions we ended up with an approximation of the FPTD for the OUP instead of an exact expression. 

%Some preliminaries on fractional Bm and its future work? 

%%%%%%%%%%%%%%%%%%%%%%%%%%%%%%%%%%%%%%
%
%	ACKNOWLEDGMENTS
%
%%%%%%%%%%%%%%%%%%%%%%%%%%%%%%%%%%%%%%

\section{Acknowledgments}

We thank Michael A. Lomholt for fruitful discussions. LL acknowledges the Knut and Alice Wallenberg foundation and the Swedish Research Council (VR), grant no. 2012-4526, for financial support. TA is grateful to VR for funding, grant no. 2014-4305.

%%%%%%%%%%%%%%%%%%%%%%%%%%%%%%%%%%%%%%
%
%	APPENDIX A: SIMULATIONS
%
%%%%%%%%%%%%%%%%%%%%%%%%%%%%%%%%%%%%%%

\appendix
\section{Simulations} \label{sec:sim}

Following \cite{gillespie} we simulate the Ornstein--Uhlenbeck process with
\begin{equation} \label{eq:gillespie}
x(t+\Delta t) = x(t)\,e^{-\Delta t} +\sqrt{1-e^{-2\Delta t}}\,\mathcal{N}(0,1)
\end{equation}
where $\mathcal{N}(0,1)$ is a normally distributed number with mean zero and variance one. This equation is on dimensionless form where we made the replacements
\begin{equation}
x\rightarrow  \sqrt{\frac{k_BT}{k}} x \ \ \ \text{and} \ \ \ t \rightarrow \frac \gamma k t,
\end{equation}
where $k_B T$ is thermal energy, $k$ the harmonic spring constant and $\gamma$ is hydrodynamic friction. The diffusion constant is $D=k_BT/\gamma$. In the simulations we varied $x_0$ and $B$ and made statistics of when $x(t)$ reached $B$ for the first time. We averaged over  $10^6-10^7$ ensembles.

It is well known that the Langevin dynamics scheme systematically overestimates the first--passage time because it can potentially miss crossings that happened within $\Delta t$. To reduce this error we used adaptive time steps that get smaller as $x(t)$ approaches  $B$. We change $\Delta t$ as follows:
\begin{enumerate}
\item Set $\Delta t=\Delta t_0$.
\item Calculate the probability that  $x(t+\Delta t)$ is above $B$ given that $x(t)$ is below $B$. That is $\omega_>(\Delta t)$.
\item If $\omega_>(\Delta t)>\epsilon$ then $\Delta t \rightarrow \Delta t/2$. Otherwise do not change $\Delta t$.
\item If at a later time  $\omega_>(\Delta t)<\epsilon$, then $\Delta t \rightarrow 2\Delta t$ with $\Delta t_0$ as upper limit. 
\end{enumerate}
In the simulations we used $\Delta t_0 =10^{-3}$ and $\epsilon =10^{-4}$.

\section{Supplementary figures}\label{sec:figs}
\begin{figure}[]
\includegraphics[width = 0.485\columnwidth]{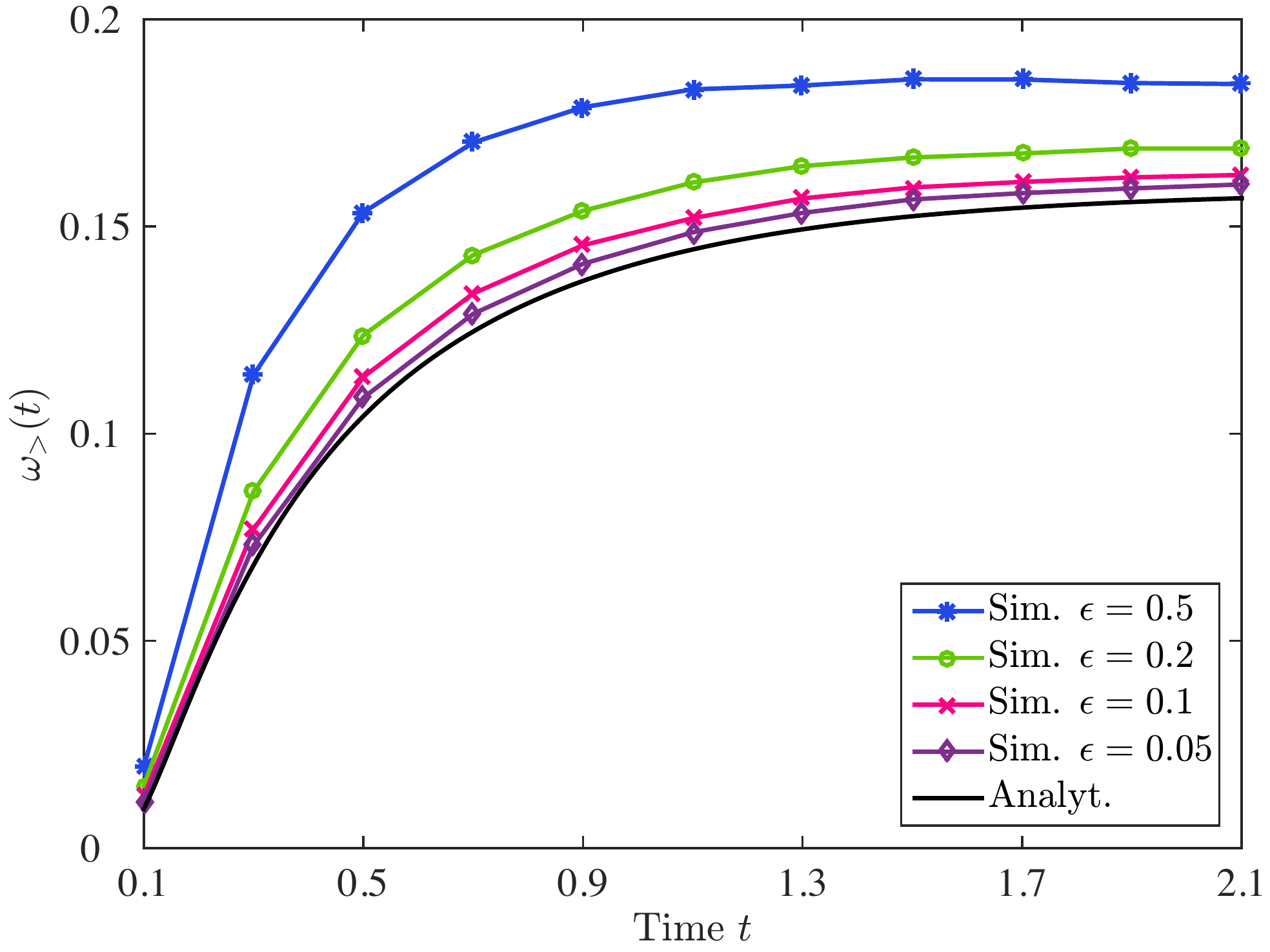}
\includegraphics[width = 0.485\columnwidth]{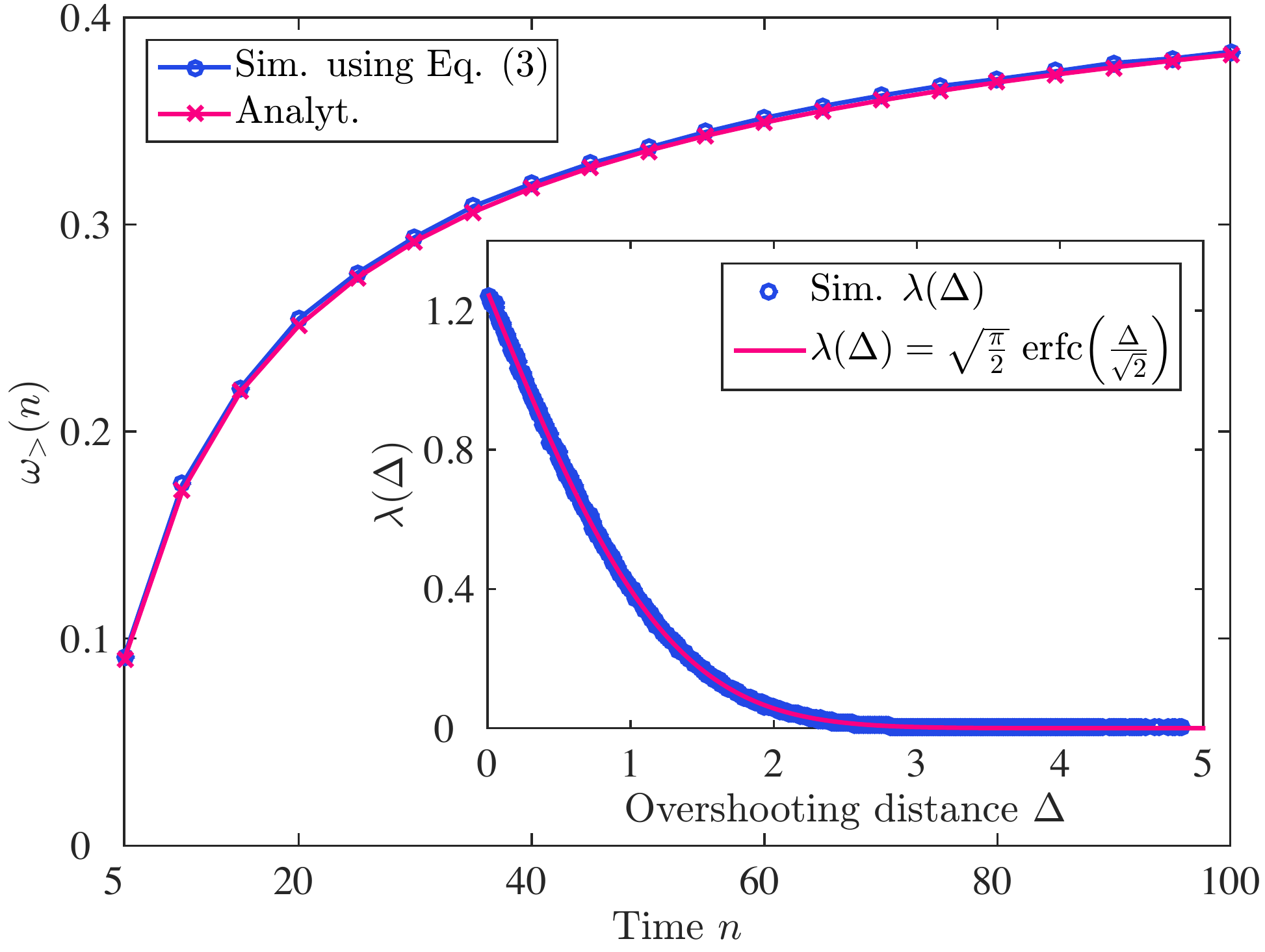}
%\includegraphics[width = 0.32\columnwidth]{omega_disc_B=3_x0=0.pdf}
%\includegraphics[width = 0.32\columnwidth]{overshoot_pdf_disc.pdf}
%\caption{(left) Convergence of the clipped process to the continuous Ornstein--Uhlenbeck process as $\epsilon\rightarrow 0$. For smaller and smaller $\epsilon$  the simulated $\omega_>(t)$ (Langevin dynamics, see Appendix  \ref{sec:sim}) gets increasingly closer to the analytical result. Here we put $B=1$ and $x_0=0$. (middle) Simulated $\omega_>(n)$ as a clipped process compared to the analytical result for the discrete unbiased Brownian walk with $B=3$ and $x_0=0$. (right) Distribution of overshooting length $\lambda(\Delta)$ when the boundary $B=0$ is crossed for the discrete Brownian walk. All simulations are averaged over $10^6$ realisations.\label{fig:clipped}
%}
\caption{(left) Convergence of the clipped process to the continuous Ornstein--Uhlenbeck process as $\epsilon\rightarrow 0$. For smaller and smaller $\epsilon$  the simulated $\omega_>(t)$ (Langevin dynamics, see Appendix  \ref{sec:sim}) gets increasingly closer to the analytical result. Here we put $B=1$ and $x_0=0$. (right) Simulated $\omega_>(n)$ as a clipped process compared to the analytical result for the discrete Brownian walk with $B=3$ and $x_0=0$. (inset) Distribution of overshooting length $\lambda(\Delta)$ when the boundary $B=0$ is crossed for the discrete Brownian walk. All simulations are averaged over $10^6$ realisations.\label{fig:clipped}
}
\end{figure}

In Fig. \ref{fig:x(t)} we illustrated how we approximate the real process $x(n)$ by the so--called clipped process. Here in Fig. \ref{fig:clipped} we show explicitly how the clipped process differs from the real $x(n)$ in for the continuous Ornstein-Uhlenbeck process and the discrete time Brownian walk. To the left in Fig. \ref{fig:clipped} we show the continuous time case. Clearly we approach the analytical curve for $\omega_>(t)$ [Eq. \eqref{eq:omega+oup}] as we let $\epsilon\to0$ [Eq. \eqref{eq:clipped+cont}] in our simulations. To the right in  Fig. \ref{fig:clipped}, we show the discrete case. We found the overshoot distribution $\lambda(\Delta)=\sqrt{\pi/2}\,\text{erfc}\left(\Delta/2\right)$ from simulations (see inset). Using  $\lambda(\Delta)$, we simulate $\omega_>(n)$ as a clipped process. We see good agreement with analytics in all aspects.

\section{Kramers escape} \label{sec:kramers}
For long times our method is consistent with Kramers escape theory. To see this we use the final value theorem and find that Eq. \eqref{eq:main+result} in the limit $s\to 0$ (long times) becomes after Laplace inversion
\begin{equation}
\rho_B(t) \simeq  \omega_>(\infty)r_+ e^{-r_-t}.
\end{equation}
Using Eq. \eqref{eq:tau+plus+norm} to eliminate $r_+$ gives
\begin{equation}
\rho_B(t) \simeq  r_-\left( 1 - \omega_>(\infty)\right) e^{-r_-t}.
\end{equation}
When $B$ is large enough we get that $1-\omega_>(\infty)\approx1$, Kramers expression \cite{kramers,hanggi}. As an example, for the harmonic potential we find that $\omega_>(\infty) \approx 0.001350$ when $B=3$ and $x_0=0$.

\section{Method of Images}\label{sec:MMI}

\begin{figure}[] 
	\centering
	\includegraphics[width=0.65\columnwidth]{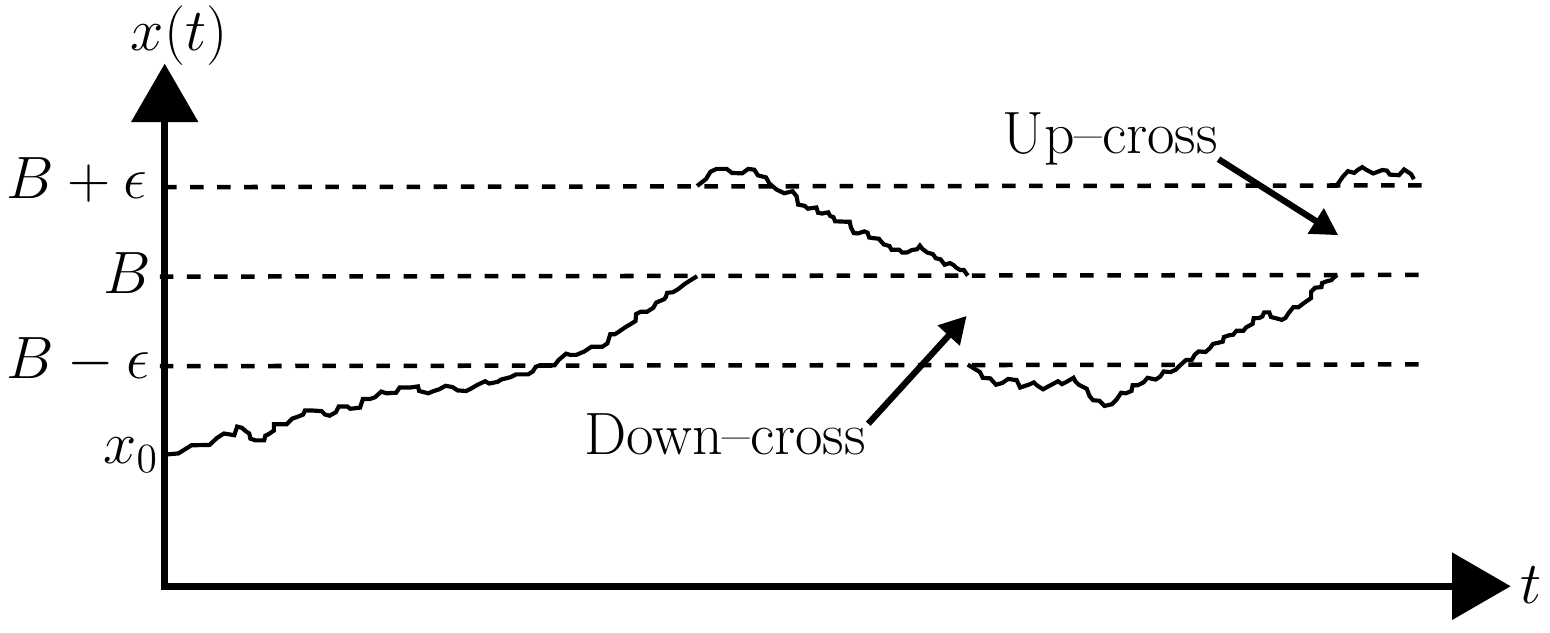}
	\caption{Separated boundary for up--crossing and down--crossing for Brownian motion.\label{fig:delta_x}}
\end{figure}

The method of images can successfully give the FPTD to a boundary for symmetric Markovian problems. Here we show that our IIA formalisms is consistent with this method for Gaussian processes. To show this, we must first find the return probabilities.

For a symmetric problem, e.g. when the absorbing boundary is a the bottom of a harmonic well, the return probability densities on either side of the boundary are equal,  $\overline\psi_+(t)=\overline\psi_-(t)\equiv \overline\psi(t)$. To find them we first realise that the Langevin equation close to the bottom of the well, say $x=0$, is similar to unbiased Brownian motion: if the force is $-x$,  the Langevin equation is $dx(t)/dt=-x+\eta(t)\approx\eta(t)$, close to $x=0$, where $\eta(t)$ is white noise. For Brownian motion we may use the following trick to find $\overline\psi(t)$.

Consider a Brownian particle that diffuses between two boundaries separated by a small distance $\epsilon$ (see Fig. \ref{fig:delta_x}). We denote an 'up--cross' by crossing $B$ from below and a 'down--cross' by crossing $B$ from above. When $x=B$ is crossed the particle jumps immediately to $B\pm\epsilon$, as explained in Sec. \ref{sec:method3}. Now, the distribution of  times between an up--cross and a down--cross is just the first--passage to a point that is a distance $\epsilon$ away, that is \cite{vankampen}
\begin{equation} \label{eq:rho+free}
\overline \psi(t) = \frac{|\epsilon|}{\sqrt{4\pi t^{3}}}\,\text{exp}\left( -\frac{\epsilon^2}{4t}  \right)
\end{equation}
with Laplace transform 
\begin{equation} \label{eq:psi+delta_x}
 \overline \psi(s) = \text{exp}\left( -\left| \epsilon \right| \sqrt{s} \, \right).
\end{equation}
If we now let $\epsilon \to 0$, up-- and  down--crossings occur to the same  boundary and therefore $ \overline \psi(s)=1$, or 
\begin{equation}
\overline \psi_{\pm}(t)=\overline \psi(t)=\delta(t).
\end{equation}
This result manifests the fractal nature of Brownian motion: if there is one  $B$--crossing at  time $t$, there will be infinitely many in the infinitesimal interval $(t,t+dt)$ \cite{blake}. 

With $\overline \psi(t)$ at hand we may derive the method of images formula. First we put $\overline \psi(s)=1$ in Eq. \eqref{eq:main+result}, which after after inversion leads to
\begin{equation}  \label{eq:rho+laplace+psi+dirac2}
\rho_{B}(t|x_0) = 2 \frac{d \omega_>(t)}{dt}.
\end{equation}
Second, since $\fptd=-d S(t)/dt$, $S(t)$ is the probability of not crossing a given boundary up to time $t$, and $S(0)=1$ and $\omega_>(0)=0$, we rewrite Eq. \eqref{eq:rho+laplace+psi+dirac2} as
\begin{equation} \label{eq:MIA4}
S(t) = 1-2\omega_>(t).
\end{equation}
For a Gaussian process with mean $\mu(t)$ and variance $\sigma^2(t)$,  $\omega_>(t)$ is 
\begin{equation} \label{eq:MIA5}
\omega_>(t) = \frac{1}{2}\text{erfc}\left( \frac{B-\mu(t)}{\sqrt{2\sigma^2(t)}} \right).
\end{equation}
Using  Eq. \eqref{eq:MIA5} in \eqref{eq:MIA4} and the relations $\text{erfc}(\bullet)=1-\text{erf}(\bullet)$ and $\text{erf}(-\bullet)=-\text{erf}(\bullet)$ leads to the method of images formula.
\begin{equation} \label{eq:MIA6}
\begin{aligned}
S(t) &= \frac{1}{2}\left[1+\text{erf}\left(\frac{B-\mu(t)}{\sqrt{2\sigma^2(t)}} \right)  \right] -\frac{1}{2}\left[1+\text{erf}\left(\frac{\mu(t)-B}{\sqrt{2\sigma^2(t)}} \right)  \right] \\
&=\int_{-\infty}^B \left[ P(x,t\,|\,x_0) - P(x,t\,|\,2B-x_0) \right]\,dx.
\end{aligned}
\end{equation}

\section{Alili's formula} \label{sec:alili}
To compare our result for the Ornstein--Uhlenbeck process to one of the best known approximations we have implemented one of the formulas from \cite{alili}. In our notation it reads
%To compare our result to one of the best known approximation of the FPTD for a harmonically trapped Brownian particle we have implemented one of the formulas given in \cite{alili}:
%
\begin{equation} \label{eq:alili+approx}
\begin{aligned}
&\fptd = \frac{e^{A/2}}{2t}\frac{H_{-A/(2t)}(-x_0/\sqrt{2})}{H_{-A/(2t)}(-B/\sqrt{2})}+\\
&\frac{e^{A/2}}{t}\sum_{k=1}^N(-1)^k \text{Re}\left(  \frac{H_{-A/(2t)-k\pi i/t}(-x_0/\sqrt{2})}{H_{-A/(2t)-k\pi i/t}(-B/\sqrt{2})} \right)
\end{aligned}
\end{equation}
where $H_{\nu}(z)$ is the Hermite function of order $\nu$. Here $A$ and $N$ are parameters that are determined based on trial and error. We found that for $t\leq10\rightarrow \{A=18.1, N=1000\}$ while for $t>10 \rightarrow \{A=7,N=1000\}$. The comparison to Langevin dynamics simulations and our IIA formula are seen in Fig. \ref{fig:one_B}. 

%%%% END TABLE TWO BOUNDARIES%%%%

%%%%%%%%%%%%%%%%%%%%%%%%%%%%%%%%%%%%%%
%
%	BIBLIOGRAPHY
%
%%%%%%%%%%%%%%%%%%%%%%%%%%%%%%%%%%%%%%

%\clearpage
%\newpage
%\bibliographystyle{unsrt} %plain 
%\bibliography{IIA} % anvŠnd \bibliography{../thesis} om thesis.bib ligger i mappen under

\end{document}